# Optimal Rate and Maximum Erasure Probability LDPC Codes in Binary Erasure Channel

H. Tavakoli
Electrical Engineering Department
K.N. Toosi University of Technology, Tehran, Iran
tavakoli@ee.kntu.ac.ir

M. Ahmadian Attari
Electrical Engineering Department
K.N. Toosi University of Technology, Tehran, Iran
m_ahmadian@kntu.ac.ir

M. Reza Peyghami
Department of Mathematics
K.N. Toosi University of Technology, Tehran, Iran
peyghami@kntu.ac.ir

***Abstract*—In this paper, we present a novel way for solving the main problem of designing the capacity approaching irregular low-density parity-check (LDPC) code ensemble over binary erasure channel (BEC).**
**The proposed method is much simpler, faster, accurate and practical than other methods. Our method does not use any relaxation or any approximate solution like previous works. Our method works and finds optimal answer for any given check node degree distribution. The proposed method was implemented and it works well in practice with polynomial time complexity. As a result, we represent some degree distributions that their rates are close to the capacity with maximum erasure probability and maximum code rate.**

## I. INTRODUCTION

One class of the most powerful error-correcting codes is Low-Density Parity-Check (LDPC) codes based on Gallager's work [1]. He introduced LDPC codes and the related iterative message-passing algorithm in 1960 in his Ph.D. thesis and then it was long forgotten until MacKay rediscovered it in 1996 [2]. After this rediscovery, searching about LDPC properties causes invention of long irregular LDPC codes with achieving and approaching to the channel capacity performance [3].

Binary Erasure Channel (BEC) is the simplest discrete channel model. Erasure channel can be used to model real world channels such as data networks. Because of this importance and simplicity, achieving and approaching to the erasure channel capacity has been studied in many articles such as [4]. Simplicity of this channel comes from Density Evolution (DE), introduced in [5] and extended in [3] and [6]. DE shows the behavior of the iterative message passing decoder in infinite number of iteration in [7-9] and the methods for code designing studied in [10], [11] and [6].

The way of code designing to approach the capacity has been remained as an open problem. Some experiments to these attempts are presented in [12-15].

The main problem of capacity approaching is its non-linear constraint that we solve it in this article by Semi-Definite Programming (SDP) approach.

Apart from our method, there are three ways for finding good degree distribution. First, using an evolutionary optimization method such as; genetic algorithm [3, Section IV.] which is not efficient and convergent. Second, optimization algorithms based on Differential Evolution, to search the answer [3, Section IV.]. These types of algorithms have infinite number of iterations which may not converge to the optimal answer. Third, a method with missing some answers with descretizing the non-linear constraint in a set of fractional values in $[0,1]$ used [7, Section V.]. So, the problem

$$\text{Min} \sum \frac{\rho_j}{j}$$

$$\text{Subject to:} \quad \sum \rho_j\big(1-\varepsilon\lambda(x)\big) > 1-x$$

changes to

$$\text{Min} \sum \frac{\rho_j}{j}$$

$$\text{Subject to:} \quad \sum \rho_j\big(1-\varepsilon\lambda(x_i)\big) > 1-x_i.$$

For a set $\{x_0, x_1, \ldots, x_N\} \subseteq (0,1]$.

The other methods for designing degree distribution are presented in [16] and [17] for finding variable degrees with using Taylor's series of $\rho^{-1}(1-x)$ in non-linear constraint $\rho(1-\varepsilon\lambda(x)) > 1-x$.

In contrast to these methods, our method is based on the exact constraint with no relaxation. By relaxation, an optimization problem the answer would be sub-optimal. With this consideration, it is certified that the answer of our SDP problem would be optimal. This approach effectively used for finding optimal codes without maximum erasure probability in [18].

The organization of the paper is as follows. In Section II, we provide the general form for the main problem of optimizing degree distribution. In Section III, we describe SDP reformulation for optimal rate problem. In Section IV, we introduce how we can optimize a code in practical numerical simulation. At last, in Section V, we illustrate our contribution with some simulation results.

## II. PROBLEM DEFINITION

Now, we analyze irregular LDPC code ensemble over BEC channel. Let G be a Tanner graph with k message bits in random with two edge polynomials specified by:

$$\rho(x) = \sum_{j=2}^{D_c} \rho_j x^{j-1} \qquad \lambda(x) = \sum_{i=2}^{D_v} \lambda_i x^{i-1} \tag{1}$$

where $D_c$ and $D_v$ are maximum check and variable node degrees, respectively. The coefficients of both polynomials represent the fraction of edges related to each variable and check nodes, i.e.,

$$\sum_{j=2}^{D_c} \rho_j = 1 \qquad \sum_{i=2}^{D_v} \lambda_i = 1 \qquad \lambda_i \geq 0, \rho_j \geq 0 \tag{2}$$

Let $\overline{d_c} = 1/\left(\sum_{j=2}^{D_c} \rho_j/j\right)$ and $\overline{d_v} = 1/\left(\sum_{i=2}^{D_v} \lambda_i/i\right)$ denote the average check and average variable nodes, respectively. It is well known that the code rate is defined as [7]:

$$R = 1 - \frac{\overline{d_v}}{\overline{d_c}} \tag{3}$$

For BEC with erasure probability $\varepsilon > 0$ , the related channel capacity is $C = 1 - \varepsilon$.The necessary and sufficient condition for achieving zero error probability, infinite number of iteration for decoding considered, which comes from DE, is

$$\lambda\left(1-\rho(1-x)\right) \leq \frac{x}{\varepsilon} \qquad \forall x \epsilon [0, \varepsilon] \tag{4}$$

Suppose fixed check node degree. In order to maximize the rate of the designing code, it is sufficient to solve the following optimization problem:

$$\text{Min } t \tag{5}$$

$$\begin{aligned} \text{Subject to:} \quad & \lambda_i \geq 0 \\ & \sum \lambda_i = 1 \\ & \sum \lambda_i \left(1-\rho(1-x)\right)^{i-1} \leq tx \quad \forall x \epsilon (0,1] \\ & t = \frac{1}{\varepsilon} \\ & t \geq 1. \end{aligned}$$

Because of an inequality exist in constraints; this problem is a semi-infinite optimization problem with infinite number of linear constraints. According to [7, Section V.], one way for solving this problem is partitioning the continuous interval set $x \in (0,1]$ to discreet set $\{x_0, x_1, \ldots, x_N\} \subseteq (0,1]$. The main disadvantage for this way is that a sub-optimal solution would be achieved. It is clear that this optimization problem is linear with respect to the cost function values $\left(\lambda_1, \lambda_2, \ldots, \lambda_{D_v}\right)$ and t.

Now, for solving the semi-infinite problem, we are going to reformulate this non-linear constraint as a Linear Matrix Inequality (LMI) and therefore we get a semidefinite reformulation for Eq. (5). It is notable that this reformulation leads to an exact solution for solving the problem instead of suboptimal solutions. So, this problem will solve by using polynomial time interior-point methods. Numerical results confirm our claim in comparison with the existence method.

## III. SDP REFORMULATION FOR THE OPTIMAL RATE PROBLEM

### *A. Problem Formulation*

In this section we reformulate the inequality constraint in optimization problem, Eq.(5), as an equivalent semidefinite programming problem in order to solve it with polynomial time interior-point methods.

Let us briefly describe our way. First, we discuss about the main constraint of the problem. In other words, the feasible region of the problem Eq.(5) contains all the vectors $\lambda$ that satisfy the following inequality:

$$P(x) = \lambda\big(1 - \rho(1 - x)\big) \leq tx \quad \forall x \epsilon (0,1] \tag{6}$$

It is clear that the function $P(x)$ is a polynomial function with degree at most $D_c D_v$. Let $q = D_c D_v$ and

$$P(x) = \sum_{j=1}^{q} p_j x^j \tag{7}$$

where $p_j = p_j\big(\lambda_1, \lambda_2, \ldots, \lambda_{D_v}, t\big)$

In the general form of a polynomial Eq.(6) we reformulate it as an LMI, i.e.,

$$P(x) \geq 0, \qquad \forall x \in \mathbb{R}. \tag{8}$$

and q=2k is an even number. It proved in [19] that semi-definite representation of infinite constraints such as Eq.(8) is semi-definite representable and its semi-definite representation is:

$$\{\lambda | P(x) \geq 0, \ \forall x \in \mathbb{R}\} = \left\{\lambda \middle| \exists\ B \in S_{+}^{k+1};\ p_l = \textstyle\sum_{i+j=l} B_{ij}, \quad \forall l: 0 \leq l \leq q = 2k\right\} \tag{9}$$

$S_{+}^{k+1}$ is the set of all positive semi-definite symmetric matrices with order k+1. It is clear that we can extend the above mentioned representation to the set $\left\{\lambda \middle| (1 + x^2)^q P\left(\frac{x^2}{1+x^2}\right) \geq 0,\ \forall x \in [0, \infty)\right\}$. In the next section, we provide a semi-definite representation for this set.

## IV. CODE OPTIMIZATION

In this section, we show that how an explicit semi-definite representation for the inequality represents in Eq.(5). In fact, we replace the infinitely many constraints of Eq. (6) by some finite LMIs and a positive semi-definite symmetric matrix. Our method, instead of ignoring some of these constraints by discretizing the interval [0, 1] to finite points as it has been presented in the literature [7], solves the problem Eq.(5) by all constraints, The following Lemma and Theorem lead us to the aim of this section.

**Lemma1**: Let $\Pi(x) = (1 + x^2)^q P\left(\frac{x^2}{1+x^2}\right) = \sum_{j=0}^{2q} \Pi_j x^j$, where $P(x)$ is defined in Eq. (11). Then we have:

$$\Pi_t = \begin{cases} \sum_{i=1}^{j} \binom{q-i+1}{j-i+1} p_{i-1} & t = 2j \\ 0 & t = 2j+1 \end{cases}. \tag{10}$$

**Proof**: We have:

$$\Pi(x) = \sum_{j=0}^{q} p_j x^{2j} (x^2 + 1)^{q-j} \tag{11}$$

By using Newton's expansion, we obtain:

$$x^{2j}(x^2 + 1)^{q-j} = x^{2j} \sum_{r=0}^{q-j} \binom{q-j}{r} x^{2r} = \sum_{r=0}^{q-j} \binom{q-j}{r} x^{2r+2j} = \binom{q-j}{0} x^{2j} + \binom{q-j}{1} x^{2+2j} + \binom{q-j}{2} x^{4+2j} + \cdots + x^{2q}$$

Therefore,

$$\Pi(x) = \sum_{j=1}^{q} \left\{p_j \binom{q-j}{0} x^{2j} + p_j \binom{q-j}{1} x^{2+2j} + p_j \binom{q-j}{2} x^{4+2j} + \cdots + p_j x^{2q}\right\}$$

This easily shows that

$$\Pi_t = \begin{cases} \sum_{i=1}^{j} \binom{q-i+1}{j-i+1} p_{i-1} & t = 2j \\ 0 & t = 2j+1 \end{cases}$$

which completes the proof of the Lemma.

**Theorem1**: Let $\Pi(x)$be defined as in Lemma 1. Then, the problem Eq.(5) is equivalent to the following semi-definite programming problem:

$$\begin{aligned} &\text{Min } t \\ &\text{Subject to:} \quad \textstyle\sum \lambda_i = 1 \\ &\qquad\qquad\quad \Pi_l = \textstyle\sum_{i+j=l} B_{ij}, \qquad 0 \leq l \leq 2q \\ &\qquad\qquad\quad B \succcurlyeq 0,\ \ 0 \leq \lambda_i \leq 1, \quad t \geq 1 \end{aligned}$$

where $\geq$ is the component-wise order on the vectors and $\succcurlyeq$ denotes the Lowner partial order on symmetric matrices that stands for positive semi-definiteness of the matrices.

**Proof**: According to the discussions of pervious section, the vector λ satisfies Eq. (7) if and only if its image by affine mapping $P(x) \rightarrow \Pi(x) = (1 + x^2)^q P\left(\frac{x^2}{1+x^2}\right)$ from $\mathbb{R}$ to $[0,1]$satisfies $\Pi(x) \geq 0$, for all $x \in \mathbb{R}$. Using Eq.(9), this equality happens if and only if there exists a symmetric positive semi-definite matrix $B = \left(B_{ij}\right)_{(q+1)\times(q+1)}$ so that it satisfies the following equations:

$$\begin{cases} \Pi_l = \sum_{i+j=l} B_{ij}, & 0 \leq l \leq 2q \\ B \succcurlyeq 0, \end{cases}$$

The proof is completed by replacing these system of linear equations and LMIs in Eq.(5) [19].

In order to illustrate these results, we provide a simple structured example to show how these results can be handled in the real problems and computer programming.

**Example1**: Suppose that we are looking for the maximum value of the parameter a so that the polynomial function $f(x) = ax^2 + bx + c$ be nonnegative in the interval [0, 1] under the condition $b = c = 1$. We apply the above-mentioned results and we first obtain the coefficients of the polynomial $\Pi(x)$ using Lemma 1 as follows:

$$\Pi(x) = (1 + x^2)^2 f\left(\frac{x^2}{1 + x^2}\right) = (a + b + c)x^4 + (b + 2c)x^2 + c = (2 + a)x^4 + 3x^2 + 1$$

Using Theorem 1, the equivalent semi-definite programming reformulation of the problem is defined as follows:

Max a

Subject to: $y_2 = 1$

$y_3 + y_5 = 0$

$y_4 + y_6 + y_8 = 3$

$y_7 + y_9 = 0$

$-y_1 + y_{10} = 2$

$y_1 = a$

$$\begin{bmatrix} y_2 & y_3 & y_4 \\ y_5 & y_6 & y_7 \\ y_8 & y_9 & y_{10} \end{bmatrix} \succcurlyeq 0$$

Using SDP softwares, such as SeDuMi and CVX, lead us to optimal solution $a = 1$, which can be verified also by using classical solution ways such as interior point method [20].

## V. SIMULATION RESULTS

Now, we present some simulation results obtained by computer numerical simulation. In this simulation, both regular and irregular parity check node degree distributions are considered.

**Example2**: For $\rho(x) = x^4$, the best variable degree distribution of degree 4 is $\lambda(x) = 0.4393x + 0.2097x^2 + 0.0536x^3 + 0.2974x^4$. The corresponding rate is $0.421$ with Capacity $= 0.44$.

**Example3**: For $\rho(x) = x^5$, the best variable degree distribution of degree 6 is $\lambda(x) = 0.4021x + 0.2137x^2 + 0.3902x^6$. The corresponding rate is $0.4922$ with Capacity $= 0.51$.

**Example4**: For $\rho(x) = 0.48555x^5 + 0.51445x^6$, the best variable degree distribution of degree 6 is $\lambda(x) = 0.4032x + 0.1512x^2 + 0.4454x^6$. The corresponding rate is Rate $= 0.5267$ with Capacity $= 0.55$.

The criteria for comparing these methods are [6]:

1-lower maximum degree

2-higher rate

3-higher erasure probability

4-lower fraction of degree-two edges

5-gap to the capacity $\delta = 1 - \mathrm{R/C}$

TableI. compares the results obtained from our method with the best results obtained in the literature.

TABLE I. COMPARING OUR NUMERICAL RESULTS VS. OTHER METHODS

| | Our result | Type-A [17] | Type-B [17] | [20] | [6.Example3.63] |
|---|---|---|---|---|---|
| ε | 0.49 | 0.48 | 0.48 | 0.5 | 0.4741 |
| $d_c$ | 5 | 5 | 5 | 6 | 8 |
| $d_v$ | 6 | 12 | 7 | 14 | 19 |
| δ | 0.0349 | 0.0389 | 0.0527 | 0.14 | 0.0493 |

As it appears, by comparing to the best results obtained until now, our way for designing good LDPC codes leads to an optimal solution in high erasure probability and maximum code rate with low fraction of degree-two edges in high erasure probability based on the above 5 criteria.

## VI. CONCLUSION

In this paper, we solved the rate optimization problem by using our reformulation. We represented a semi-definite problem without any relaxation or simplification. Simulation results show that our results in most cases are better than the best reported results in both rate maximization and maximum erasure probability in the literature.

## ACKNOWLEDGEMENT

This work was partially supported by ITRC of IRAN.